\documentstyle[12pt]{article}

\begin{document}

\def\fluxd{\hbox{cm${}^{-2}$s${}^{-1}$MeV${}^{-1}$}}
\def\fluxt{\hbox{cm${}^{-2}$s${}^{-1}$}}
\def\li#1{\hbox{${}^{#1}$Li}}
\def\he#1{\hbox{$^{#1}{\rm He}$}}
\def\bor1#1{\hbox{${}^{1#1}{\rm B}$}}
\def\car1#1{\hbox{${}^{1#1}{\rm C}$}}
\def\oxy1#1{\hbox{${}^{1#1}{\rm O}$}}
\def\dbbe{\hbox{$\partial_t{\rm B}/\partial_t{\rm Be}$}}
\def\dlibe{\hbox{$\partial_t{\rm Li}/\partial_t{\rm Be}$}}
\def\beq{\begin{equation}}
\def\eeq{\end{equation}}

\rightline{FERMILAB--Pub--94/-A}
\rightline{astro-ph/9407099}
\rightline{July, 1994}

\vspace{.2in}

\begin{center}

{\bf IMPLICATIONS OF A HIGH POPULATION II B/Be RATIO}

B.D. Fields$^{1}$, K.A. Olive$^{2}$, and D.N. Schramm$^{1,3}$ \\

$^{1}${\it The University of Chicago \\
5640 S. Ellis Ave., Chicago, IL  60637-1433}

$^{2}${School of Physics and Astronomy \\
The University of Minnesota, Minneapolis, MN 55055}

$^{3}${NASA/Fermilab Astrophysics Center \\
Fermi National Accelerator Laboratory, Batavia, IL  60510-0500}

\end{center}

\vspace{.3in}

\centerline{\bf abstract}

The observed B/Be ratio in extreme Pop II stars
has been interpreted as evidence
of Be and B synthesis by early galactic cosmic rays.
However, a recent reanalysis of the boron abundance
in the Pop II halo star HD140283 suggests that B/H may be
larger than previously reported, by as much as a factor of 4.
This would yield a B/Be
ratio lying in the range $14 \la {\rm B/Be} \la 50$.
The possibility of a high Pop II B/Be ratio
stresses the importance of the
upper limit to the B/Be
ratio arising from cosmic ray production.
It is found that the limit to cosmic ray-produced B/Be depends upon
the assumed cosmic ray spectrum.
For any Pop II
comic ray spectrum that is a single power law in
either total energy per nucleon or in momentum
%(both of which are consistent, for a particular spectral index,
%with the present observed flux)
the B/Be ratio is constrained to lie in the range
$7.6 \la$ B/Be $\la 14$.
Thus, if the new B/Be ratio is correct,
it requires either a bimodal cosmic ray flux with a large
low energy component, or for
another B source, possibly the proposed $\nu$-process
in supernovae, either of which may be helpful in explaining the
observed \bor11/\bor10 ratio.
Finally, it is noted that the boron reanalysis
highlights the uncertainty in our knowledge of the B/Be
ratio, and the need for additional data on Be and B abundances.

\end{abstract}

In the last few years, new observations of  Population II halo stars
have led to the detection of B (Duncan et al.\ 1992 (DLL);
Edvardsson et al.\ 1994)
and Be (Rebolo et al.\ 1988; Ryan et al.\ 1990, 1992;
Gilmore et al.\ 1992a, 1992b; Boesgaard \& King 1993). It is
commonly believed
(Reeves et al.\ 1970;
Meneguzzi et al.\ 1971 (MAR);
Reeves et al.\ 1973;
Walker et al.\ 1985; Steigman \& Walker 1992 (SW);
Prantzos et al.\ 1993 (PCV); Walker et al.\ 1993 (WSSOF);
Steigman et al.\ 1993 (SFOSW); Fields et al.\  1994 (FOS))
that these elements have their origin
in early cosmic ray activity. Spallation of carbon, nitrogen and oxygen
by protons and $\alpha$ nuclei can for the most part account
for the observed abundances of B and Be.
Early cosmic rays can also produce
some of the observed \li7 as well as
all of the now observed \li6 (Smith et al.\ 1992;
Hobbs \& Thorburn 1994),
in part by spallation but predominantly via the accompanying
$\alpha + \alpha$ fusion.

A comparison of observed abundance ratios and their theoretical
predictions is a good test of models of galactic cosmic ray
nucleosynthesis (SW; PCV; WSSOF; SFOSW; FOS) and galactic
chemical evolution (PCV);
it may have implications for big bang nucleosynthesis as well
(WSSOF; Olive \& Schramm 1993). The ratios of interest are
\li6/\li7, Li/Be, B/Be, and potentially \bor1{1}/\bor10.
In the case of \li6/\li7
where the theoretical prediction of about 0.9 (from
cosmic-ray nucleosynthesis) is robust, the observation
of \li6 (Smith et al.\ 1992;
Hobbs \& Thorburn 1994) is a good indication
that Li is not strongly depleted in stars (at least not by
nuclear burning (Brown \& Schramm 1988; Deliyannis et al.\ 1989)).
Though caution is still warranted due to the current paucity of data,
the \li6/\li7 ratio found by both groups
is consistent with standard models of cosmic-ray and big bang
nucleosynthesis and standard stellar models
which have minimal Li depletion (SFOSW).
The Li/Be ratio which can be used to probe the compatibility
between cosmic-ray and big bang nucleosynthesis (WSSOF; Olive \& Schramm
1992) is much more model-dependent (FOS).
There is as yet not data on the \bor11/\bor10 ratio in Pop II objects, but
such data would be very interesting, as this ratio is anomalous even
in Pop I objects, a point we will return to below.
Finally,
the B/Be ratio, like the \li6/\li7 ratio, is largely independent of
cosmic ray models
(WSSOF; FOS) and so is an excellent test of these models.

While there are many observations giving the Be abundance in halo
stars (Rebolo et al.\ 1988; Ryan et al.\ 1990, 1992;
Gilmore et al.\ 1992a, 1992b; Boesgaard \& King 1993),
there is data on B for only three stars
(DLL; Edvardsson et al.\ 1994) since
the B lines reside well into the ultraviolet and thus require
satellite observation.  The data are summarized in Table 1.
In the table we show the observed abundances
of Be and B as well as Fe for the three halo stars.
For each particular Be measurement we list the Fe abundance
used for that measurement.  Note that both the Be and the Fe abundances
for each star vary among the different measurements.

\begin{table*}
\caption{Observed Pop II
abundances of Be and B}
\begin{tabular}{lcccccc}
\hline
{\sc Star}& [Fe/H] & [Be] & [B] & {\sc LTE} B/Be&
{\sc NLTE} B/Be & {\sc Source}* \\
\hline
HD 19445 & -2.1 &  $-0.14 \pm 0.1 $ &
 $0.4 \pm 0.2$  & 3.4 $\pm $ 1.8 & $\sim$8 & BK\\
HD 140283 & -2.7 &  $-1.25 \pm 0.4$
 & $-0.16 \pm 0.14$& 12 $\pm$ 12 & 34 -- 50 & Ry\\
HD 140283 & -2.8 &  $-0.97 \pm 0.25$
 & $-0.16 \pm 0.14$&$7 \pm 5$ &23 -- 33 &G\\
HD 140283 & -2.7 &  $-0.78 \pm 0.14$
 & $-0.16 \pm 0.14$&$5 \pm 2$ &14 -- 21 &BK\\
HD 140283 & -2.5 &  $<-0.90$
 & $-0.16 \pm 0.14$&$>7$ &$>$ 21 -- 30 &M\\
HD 201891 & -1.3 &   $0.65 \pm 0.1$
  & $1.7 \pm 0.4$ & 10 $\pm$ 10 &$\sim$14.5 &Re, BK\\
\hline
\end{tabular}

\footnotesize
${}^{\rm *}$BK =  Boesgaard \& King 1993;
Ry = Ryan et al.\ 1990, 1992;
G = Gilmore et al.\ 1992a, 1992b;

M = Molaro et al.\ 1993;
Re = Rebolo et al.\ 1988;
\normalsize
\end{table*}

In the case of HD 140283, there are several independent
observations of Be and two observations of B.
In the table, we show the quoted value of [Be]. For [B],
we have averaged the two measurements
and, to minimize systematics,
we have adjusted the B abundance quoted in Edvardsson et al.\ (1994)
by assuming stellar parameters (temperature and surface gravity)
as in DLL. To obtain the B/Be ratios we
use the average B abundance
and we have adjusted the Be abundances in each case to also match the DLL
stellar parameters.
For HD 19445 we note that there
are in addition upper limits of [Be] $< 0.3$ (Rebolo et al.\ 1988) and
$<-0.3$ (Ryan et al.\ 1990) giving B/Be $> 1.3$ and $> 5$ respectively
which have not been included in the table and for HD 201891
the values of [Be] and
B/Be given represent an average of the two published measurements.

One should be aware that most observational determinations have
been made using different sets of parameters
in their stellar atmosphere models.
Though one can ascribe some uncertainty to chosen values of these parameters,
it is not always clear to what extent these systematic
errors have been incorporated into the quoted so-called ``statistical error,''
and different authors make divergent assumptions on the uncertainty of
their assumed stellar parameters. Thus some care is warranted in
using this data.
Since systematic errors due to assumed model parameters, etc., are probably
not distributed in a gaussian manner, nor will they be decreased
with the square root of the number of observations, one cannot
reliably apply standard statistical techniques.
(Perhaps future observational papers might consider separating the
systematic portion of the stated error from the statistical portion as
is now being done in many nuclear and particle physics papers.)

As one can see from the B/Be ratios in Table 1,
some of the LTE ratios are in
agreement with standard cosmic-ray nucleosynthesis model  predictions
(B/Be $\simeq 12-14$), but most  of them are on the low side of the prediction.
For example, the overall average in the case of HD 140283, gives
B/Be = $6 \pm 2$. (Though recall the caveat regarding systematic errors).
Thus effort has been concentrated for the most part
in determining  how low the B/Be can be made within the context of
cosmic-ray nucleosynthesis.  In WSSOF, it was argued on the basis of
spallation cross-sections that the extreme lower limit is B/Be $\ga 7$.
Both WSSOF and PCV have noted that a low Pop II B/Be ratio (between 7 and 10)
would imply a flatter cosmic ray spectrum in the early Galaxy,
which is suggested to have arisen from  stronger cosmic ray  confinement.

Recently, Kiselman (1994) has performed a reanalysis of the
inferred B in HD 140283 from the DLL data.
In the original analysis of DLL, abundances
based on the B{\sc i} and Be{\sc ii}
spectral lines were extracted using the assumption
of local thermodynamic equilibrium (LTE).  The beryllium abundance
is believed to be relatively insensitive to this approximation.
It was recognized by DLL that a non-LTE (NLTE)
analysis could  be a potentially important correction to the boron abundance.
Kiselman (1994) has in great detail attempted to account for the NLTE
correction for the specific case of HD 140283. Indeed he found an
overall upward correction to the boron abundance
of 0.56 dex or a factor $\ga$ 3.
To test the reliability of his results, Kiselman perturbed
his model and estimates that a reasonable NLTE correction
to the boron abundance of
HD 140283 should lie between 0.46 and 0.62 dex.
Recently the DLL measurement of B in HD 140283
has been confirmed by Edvardsson et al. (1994). Within errors,
there is very good agreement in the LTE abundances. Edvardsson et al.\ (1994)
argue for a similar NLTE correction to their derived abundance.
In Table 1,
we also give Kiselman's
corrected B/Be ratio for the range 0.46-0.62 dex.
 The weighted average of the three
positive observations of Be (again corrected for differing surface
gravities) is [Be] = -0.93 $\pm$ .12, giving B/Be = 6 $\pm $ 2.
After the Kiselman correction we find that B/Be = 17 - 25, using the
central value of Be.  The range here corresponds to the range in
the correction factor, not to statistical errors. The correction factors for
HD 19445 and HD 201891 were obtained from Kiselman (private communication).

The possibility of a high Pop II B/Be ratio can have interesting
consequences.  WSSOF compute an upper bound
of B/Be $\la 17$ for cosmic ray production in Pop II.
However, whereas the WSSOF lower limit
to B/Be is model-independent, their upper limit is not,
as it was calculated in their ``zeroth order'' model.
The question we ask here is, what is
the true, model-independent upper bound to the B/Be
ratio arising from cosmic rays?  In this note we compute
the range of B/Be produced in various models of cosmic ray
synthesis of LiBeB,
and we discuss the implications
for alternate means of boron production.

There are several factors which affect
the maximum B/Be ratio that Pop II cosmic rays can produce.
Ultimately, the predicted ratios are controlled by
(well-measured) nuclear physics, in the guise of
spallation/fusion cross sections.  The model-dependent
feature one may adjust
is the Pop II cosmic ray flux spectrum,
which one must decide how to parameterize.
Given a choice of flux, its LiBeB yields are
constrained to be consistent with
the observed Pop II LiBeB abundances and ratios.
To determine the maximum B/Be, then,
the game is to choose a range  of admissible Pop II
cosmic ray spectra, and then to convolve it with the
cross sections find the highest B/Be ratio these
spectra can produce without violating observational constraints.

The rate of LiBeB production by cosmic rays is given for each process
by the usual rate equation (see FOS for details on our recent analysis):
the product of target abundances with an integral of the cosmic
ray flux times the cross section for the reaction and a factor
accounting for the probability of the LiBeB being stopped in the Galaxy
before escape.
The lower bound for the integral is
the threshold energy $T^0$ for each spallation/fusion process.
The thresholds are determined by $Q$ values for the reactions.
Here, the most important fact about these
thresholds is that for all spallation reactions
$T_{\rm B}^{0} < T_{\rm Be}^{0}$, i.e.\ {\it the threshold for boron production
is lower than that for beryllium production}.
Thus all of the flux
in the energy range $T_{\rm B}^{0} \le T \le T_{\rm Be}^{0}$
(in our case, 3.13 MeV $\le T \le$ 17.5 MeV)
will produce only boron.  Clearly,
one can make B/Be arbitrarily high
by tuning the low-energy cosmic ray flux to exploit this
difference in thresholds.
Note as well that
for boron isotopic production, we have
$T_{11}^{0} < T_{10}^{0}$, and so a large low energy flux will also
have the effect of increasing the \bor11/\bor10 ratio.

The cosmic ray spectrum $\phi$ is propagated
(in energy space) from a source spectrum $q$
which one must specify.
Today we observe the propagated flux from contemporary sources,
{}from which we can infer a source spectrum.
However,
observations of the present spectrum
are limited by solar modulation to include only cosmic rays
with kinetic energy per nucleon $T \ga 100$ MeV/nucl.
The observed spectrum is consistent, over this range,
with a source law taking the form of
a single power law, either in momentum per nucleon,
$q(p) \propto p^{-\gamma}$, or
in total energy per nucleon, $q(T) \propto (T+m_p)^{-\gamma}$.
The observed galactic cosmic ray flux today corresponds to such a flux,
with a spectral index of $\sim 2.7$.   The present cosmic ray
confinement is characterized by a pathlength $\Lambda$
which varies in energy around $\sim 10 \, {\rm g/cm}^2$.

We do not know directly how the cosmic ray flux behaves
at low energy ($\la 100$ MeV/nucl);
this is an unfortunate state of affairs,
as the B/Be ratio is very sensitive to
the details of the flux at precisely this energy range.
We will in this paper assume  that we may extrapolate
the cosmic ray flux from the measured high-energy region
down to the low energy regime.  We will for the moment
also assume that the low energy flux obtained through this extrapolation
is the {\em only} low energy component.
We remind the reader, however, that while the data we are extrapolating from
measures the present, Pop I cosmic ray flux, we wish to model its
behavior in the Pop II epoch.
As several authors have pointed out, in this epoch,
the flux parameters, namely
the spectral
index and escape pathlength, are not well constrained,
and indeed could have been different than those today.
We therefore will allow for these parameters to vary within
physically allowable ranges as done in FOS.

In choosing the allowed flux parameters we
consider the possibility that a high B/Be might change the outlook on
the behavior of early cosmic ray confinement.  Before the
Kiselman (1994) result, it was
argued that
larger early confinement was needed to reproduce a
low B/Be.  Now we consider the opposite case, and so
this motivation for a larger confinement weakens.
Thus, we
have allowed the escape pathlength to vary over the range
$10\ {\rm g/cm}^2 \le \Lambda \le 1000\ {\rm g/cm}^2$, which
encompasses the present values and extends up
to values at which nuclear inelastic losses dominate the escape losses.

We have calculated LiBeB production rates using cosmic ray
fluxes propagated from different source spectra that are either power laws
in momentum or in total energy.
For each source type (i.e.\ power
law in momentum or in total energy) we plot the ratio of production
rates, \dlibe, as well as
\dbbe.  Our results appear in Figs.\ 1 and 2.
As noted in FOS, lacking a model for the
galactic chemical evolution, one can only calculate the
ratio of LiBeB production rates
rather than the actual abundance ratios, e.g. B/Be or Li/Be.
However, FOS note that evolutionary processes will serve to
make \dlibe\ a lower bound for the true Li/Be ratio,
while evolutionary effects are unimportant in the B/Be ratio
which can be identified with the \dbbe\ ratio.

Note that the case of the
momentum source spectrum, (Fig.\ 1) the \dbbe\
ratio does rise with increasingly
steep spectra.  This is expected:  a featureless power law in
momentum
has a lot of power at low energies, and so the
\dbbe\ ratio should be sensitive to the spectral index (though the steepness
of the source law is greatly softened at low energies by
ionization losses included in the propagation).  However,
while the \dbbe\ ratio linearly increases with the spectral index,
the \dlibe\ ratio increases exponentially.  But a large Li/Be
ratio is constrained by the observational data.  If we demand that
the cosmic rays do not wash out the Spite plateau, then
we may very generously insist that
$({\rm Li/Be})_{\rm CR} < ({\rm Li/Be})_{\rm OBS} \simeq 1000$.
Bearing in mind
that the \dlibe\ ratio {\it underestimates} the Li/Be, we see that
the spectral index is strongly constrained.  Even for a high confinement,
the steepest allowed spectrum has $\gamma \la 3.3$.   In this range, for
all confinement parameters, $\dbbe \simeq {\rm B/Be} \la 14$.  A momentum
source that is not otherwise enhanced at low energies cannot produce
B/Be in the range of the Kiselman results.

\begin{figure}[htbp]
\picplace{10cm}
\caption{Ratios of LiBeB production rates for a
source spectrum $q(p) \propto p^{-\gamma}$.
Plotted as a
function of spectral index $\gamma$.  For both plots we use
CNO abundances
[C/H] = [N/H] = [Fe/H] = [O/H] - 0.5 = -2.5, and
\he4/H = 0.08.}

{\bf a} The \dbbe\ ratio; the solid curve is for
$\Lambda = 10 {\rm g/cm}^2$,
the broken curve is for
$\Lambda = 1000 {\rm g/cm}^2$. Note the very restricted,
linear scale in the ordinate, showing the insensitivity of
\dbbe\ to the spectral index.  \par
{\bf b} As in {\bf a}, for the \dlibe\ ratio.
Here we see that \dlibe\ is exponentially sensitive to the
spectral index, in contrast to the results of plot {\bf a}.
As discussed in the text, the observational
constraint Li/Be $\ll 1000$ implies that $\dbbe\ \simeq{\rm B/Be} \la 14$.

\end{figure}

Similar results for a source spectrum in total energy are shown in Fig.\ 2.
Note here that there is much less of a problem with the \dlibe\
ratio.  However, this spectrum is doomed to fail
to produce high B/Be.  Because the source is a power law in total energy,
$q \sim (T+m_p)^{-\gamma}$,
the nucleon rest mass $m_p$ introduces a low-energy cutoff
which keeps the flux spectrum finite and sets
the scale for the peak in the propagated flux to be around $m_p$,
far above the tens of MeV at which one requires a large flux to fit B/Be.
This effect is seen in the flatness of the \dbbe\ curve in
Fig.\ 2.
One expects a spectral index around $\gamma = 2 - 3$, and certainly
$\gamma < 5$, which gives B/Be $\la$ 12.  However, to get a feel for the
maximum possible B/Be, we arbitrarily allow the spectral index to increase
until the \dlibe\ constraint is reached.
Even in this poorly motivated case, we have $\dbbe\ \la$ 14,
the same constraint as for the momentum spectrum.   Thus, for
either spectral type, we have an upper bound of
\beq
\dbbe\ \simeq {\rm B/Be} \la 14 ,
\eeq
a limit which is independent of the choice of confinement parameter
$\Lambda$ and allows for variation in spectral index.

\begin{figure}[htbp]
\picplace{10cm}
\caption{Ratios of LiBeB production rates.
Calculated as in Fig.\ 1, for a
source spectrum $q(T) \propto(T+m_p)^{-\gamma}$. }

{\bf a} The \dbbe\ ratio; note the larger range in $\gamma$ compared
to that of Fig.\ 1, and the even slower dependence of \dbbe\ on
the spectral index. \par
{\bf b} The \dlibe\ ratio.  Here again \dlibe\ is sensitive to
$\gamma$, but less so than for a source spectrum in momentum (Fig.\ 1).
Note also that Li/Be $\ll 1000$ again gives $\dbbe\ \simeq{\rm B/Be} \la 14$.
\end{figure}

If the NLTE correction to the B abundance in HD 140283 is correct,
then for this star B/Be $\ga 14$ (see Table 1) and more likely
to be even higher, thus
a single power law cosmic ray flux underestimates the observed
B/Be ratio.  We must therefore conclude that either (1) the cosmic ray flux is
not well described by a single power law; (2) there has been significant
stellar depletion in Pop II, which would preferentially destroy Be relative
to B because of the difference in the coulomb barriers; or (3) that
something other than or in addition to cosmic rays produce the observed ratio.
We will address point (1), suggesting a possible non-power law spectrum.
Regarding point (2), as it is argued in SFOSW, we
do not expect significant depletion in these stars, as is indicated by the
positive identification of \li6 in halo stars.
Thus we will not consider this line of reasoning further.
As for point (3), we note
that no proposed source for Be (and for \li6) other than cosmic rays has stood
the test of time, and thus lacking an alternative we will continue to
assume that these nuclei do arise from cosmic rays processes. We will
consider the possibility of additional sources to the boron abundance.

If we take the observed LTE B/Be ratio to be accurate (i.e.\ we assume
that B and Be are undepleted),
and we assume that cosmic rays (with a single power law spectrum)
produced the Be (and inevitably
some B as well), then the import of the
Kiselman (1994)
NLTE calculation is that another source of boron is needed.  As mentioned
above one possibility frequently
discussed is the superposition of a low energy component to the cosmic ray
flux.
Such a low-energy component to the cosmic rays is not directly
observable.
However, introduction of a low-energy component to the cosmic ray flux
allows additional tuning of LiBeB production beyond the above considerations of
adjusting the cosmic ray source type, or confinement.

Long before the recent Kiselman (1994) analysis, there has been
another good reason for an additional source of B, namely the boron isotopic
ratio.  It is well known that standard cosmic-ray nucleosynthesis
models predict (MAR) a value \bor11/\bor10  $\simeq$ 2.5, whereas
the observed ratio
(Cameron 1983; Anders \& Grevesse 1989)
is very close to 4.
Interestingly, the same low energy flux that will make a high
B/Be ratio will also make a large \bor11/\bor10 ratio.
Indeed, this point has been noted
by MAR as well as in subsequent cosmic ray nucleosynthesis calculations.
MAR first suggested that the cosmic rays might have a
low-energy component which
could fix the (Pop I) boron isotopic problem, and possibly the
Pop I lithium isotopic ratio as well.  Authors since then have
followed this lead in trying to reproduce the solar ratios of
B and Li, and have been moderately successful in doing so,
the most recent model being that of WMV.
The low energy
particles were proposed to be similar to those seen in solar flares, which
indeed have steep spectra.  MAR and subsequent authors have
modeled this component with a power law in kinetic energy,
with indices between 3 and 7.
PCV also have some discussion of Pop II synthesis of LiBeB by including
a low energy spectral component.
They find that while addition of this component allows for
a felicitous \bor11/\bor10 ratio, the flare component also
leads to Li overproduction at low metallicities.  PCV concluded
that such a fix to the \bor11/\bor10 ratio problem, could only
be implemented during the disk phase of the galaxy.
As such, it can not account for a high B/Be ratio in halo stars.

WMV and earlier works have calculated the LiBeB yields for the case
in which the low-energy flare component dominates the production.
For this case of a LiBeB synthesis purely by flares
(in a Pop I environment) , WMV find
that such a flux does not reproduce elemental or isotopic ratios.
In particular, they find that B/Be $\ga 100$ for a flare spectrum
with index $\beta \ga 5$, and they find that in all cases \bor11/\bor10 $\ga
5$.
They also find that Li/B $\ga$ 80, an overproduction that only
becomes exacerbated in a Pop II environment.
That these ratios fit the data poorly is indication that
a flare spectrum alone cannot dominate the LiBeB production in a Pop I
environment.  Furthermore, since the Be and B production is insensitive
to galactic evolution, these conclusions hold for a Pop II environment as
well.

While we cannot observe  a low-energy flux directly, there are two
indirect observational constraints and signatures that
have been suggested.  One is its ionization of the ISM, which
MAR employ as a constraint on the low-energy flux.  Too many
cosmic ray particles would ionize the ISM beyond the observed limits.
Also, a low energy flux creates a
distinctive $\gamma$-ray spectrum.  These $\gamma$-rays are produced
by inelastic collisions with CNO nuclei that leave the CNO in an
excited state.  The de-excitation of these states leaves a signature
of distinctive lines.
Until recently these lines, predominantly
{}from the 4.44 MeV state of $\car12^*$, and at 6.13 MeV from
$\oxy16^*$, have remained unobserved.
However, the
COMPTEL group on the Gamma Ray Observatory (Bloemen et al.\ 1994)
have observed the Orion complex at 0.75-30 MeV and report a detection
of gamma ray emission in excess of background
in the 3-7 MeV range (and only in this range).

Bloemen et al.\ (1994) report a
flux of $(1.01 \pm 0.15) \times 10^{-4} \fluxt$ (3-7 {\rm MeV}).
This is to be compared with the
calculations of Meneguzzi \& Reeves (1975), applied
to Orion, for which
one expects a flux at the $\car12^*$ peak of
$\phi_\gamma \simeq (2.5-5) \times 10^{-7} \fluxd$ for
a flare-type spectrum (and significantly less for a
spectrum from a single component momentum or total energy source).
Bloemen et al.\ (1994) suggest
that an enhancement in the low-energy cosmic
ray proton flux sufficient to match the observation leads to
a large rate for the ionization if the ISM.  Consequently Bloemen et al
argue that these $\gamma$-rays are not from energetic protons
on interstellar C and O
but instead from an enhanced component of low-energy cosmic ray C and O
on ISM hydrogen.  Recently, attempts to incorporate this new gamma-ray
observation into galactic cosmic-ray nucleosynthesis models have been presented
(Cass\'{e} et al. 1994; Reeves 1994). These
models may also predict a higher than standard B/Be ratio.  In the remaining
discussion, we consider other alternatives which predict a high B/Be ratio.

Because of the difficulty in producing the observed isotopic ratio of
\bor11/\bor10 in standard cosmic-ray nucleosynthesis models,
it has been suggested that alternative astrophysical sites
for the production of \bor11 must be found.
One such site is at the shock front of type II supernovae, as suggested by
Dearborn et al.\ (1989):  when the shock hits the
hydrogen envelope, it burns the ambient $^3$He and $^4$He producing $^7$Be.
Some of the resulting
$^7$Be combines with alpha particles to produce $^{11}$C which decays to
\bor11.  They
noted that significant \bor11 production might take place.
Subsequent calculations (Brown et al.\ 1991) have shown that these hydrodynamic
processes were not sufficient producers of these light elements
for currently preferred parameter values.

A potentially more important source for \bor11 production has been
found to result from neutrino
induced nucleosynthesis in type II supernovae (Woosley et al.\ 1990).
The inelastic
scattering of  neutrinos leads to unstable excited states which
decay by p,n or $\alpha$ emission.  These processes were included
in supernova nucleosynthesis calculations by Woosley et al.\ (1990)
where it was found that considerable \bor11 production can result as the
flux of neutrinos passes through the He, C, and Si shells of the
stellar envelope, primarily by neutrino
spallation of $^{12}$C.  The dominant product is \bor11 since it is favored
for $\nu$-spallation to knock out a single nucleon.
In addition, some synthesis of \li7 and \bor10 takes
place by this process but the production rate seems quite low.
This process is attractive as it naturally creates \bor11 without much
\bor10, and so provides the needed source of \bor11 to augment GCR
production and so reproduce the \bor11/\bor10 ratio.

Indeed the \bor11 yields from these processes
(Woosley et al.\ 1993; Timmes et al.\ 1993) were incorporated in a chemical
evolution model (Olive et al.\ 1994).
Respecting the overall constraints imposed by the LiBeB observations
in halo stars, they
were able to obtain a solar isotopic ratio
$\bor11/\bor10 \simeq 4$.  Using the boron isotopic ratio to normalize
the $\nu$-process yields, they showed
 that neutrino process nucleosynthesis
leads to a relatively model independent prediction that the $B/Be$
elemental ratio is large ($>$ 50) at low metallicities ($[Fe/H] < -3.0$),
assuming still that $Be$ is produced as a secondary element
 as is the case in the conventional
scenario of galactic cosmic-ray nucleosynthesis.
(Despite earlier conjectures (Malaney 1992), $^9$Be is not
significantly produced by the $\nu$-process).
In particular, at the metalicity corresponding to that of
HD 140283, [Fe/H] $\simeq$ -2.6, Olive et al.\ (1994) predicted that the
B/Be ratio should be close to 40.  Though still on the high
side, this is in overall good agreement
with the NLTE corrected values shown in Table 1.

To summarize our results:
the Kiselman (1994) analysis of the B abundance in HD 140283
suggests that in this star the B/Be ratio is potentially
 higher than can be accounted for by cosmic ray nucleosynthesis
with a single power law source spectrum.  This is best understood as
arising from a overabundance of boron.
If indeed the boron is high, then it must have a source that was active
in the Pop II epoch, either
low-energy cosmic rays in the early Galaxy, or an alternative, non-cosmic ray
process.  The former might be suggested by the data of Bloemen et al.\ (1994),
while the latter has a promising candidate in the $\nu$-process.
These two alternatives should be distinguishable by getting more B/Be ratios,
particularly in extremely metal deficient ([Fe/H] $\la 3$) stars,
for which the $\nu$-process should be dominant and hence the B/Be
should be much larger than in HD 140283 (Olive et al.\ 1994).

The NLTE reanalysis of the boron abundance also
underscores the difficulty
of Pop II Be and B abundance measurements.
Clearly there is a need for continued scrutiny of these
abundances, as well as for further boron data in more stellar
environments, some presumably
not having the same NLTE effects and so amenable to a test of the possibility
of high B/Be.

\bigskip
%\begin{acknowledgements}
We thank Doug Duncan, Lew Hobbs, Ruth Peterson, Julie Thorburn, Frank Timmes,
and Jim Truran
for helpful discussions.
The work of (BDF) and (DNS) was supported in part by the
DOE (at Chicago and Fermilab) and by the NASA through
NAGW-2381 (at Fermilab) and a GSRP fellowship at Chicago.
The work of (KAO) was supported
in part by  DOE grant DE-FG02-94ER-40823.
and a Presidential Young
Investigator Award.
%\end{acknowledgements}

\end{document}